\documentstyle[12pt]{article}
\newcommand{\mbo}{(X,\:{\cal A},\:\mu)}
\newcommand{\calx}{\int\limits_X}

\newcommand{\eli}{L^1\mbo}
\newcommand{\sumr}{\sum\limits_{i=0}^{r^n-\,1}}
\newcommand{\sumi}{\sum\limits_{j=0}^{r\,-\,1}}
\newcommand{\suki}{\sum\limits_{k=1}^{\infty}}
\newcommand{\gran}{\lim\limits_{n\to\infty}}
\newcommand{\be}{\begin{equation}}
\newcommand{\ee}{\end{equation}}
\newcommand{\con}{\frac{1}{\mu(X)}\cdot{\bf 1}}
\newcommand{\liti}{\lim\limits_{t\to\infty}}
\newcommand{\calh}{{\cal H}}
\newcommand{\ca}{{\cal A}}
\newcommand{\gab}{g_{\alpha\beta}}
\newcommand{\gba}{g_{\beta\alpha}}
\newcommand{\dia}{\mbox{diag} }
\newcommand{\aaa}{A_\alpha}
\newcommand{\ra}{\rho_\alpha}
\newcommand{\la}{\Lambda_\alpha}
\newcommand{\vni}{\vec{n}_i}
\newcommand{\kala}{\frac{4}{3}\kappa\alpha^2}
\newcommand{\ekl}{e^{-\kala t}}
\begin{document}
\title{Completely Mixing Quantum Open Systems and Quantum Fractals}
\author{Ph. Blanchard\footnote{corresponding author, fax: 49 521 106-2961,
e-mail: blanchard@physik.uni-bielefeld.de}, A. Jadczyk\\
and R. Olkiewicz\footnote{permanent address: Institute of Theoretical
Physics, University of Wroc{\l}aw, PL-50-204 Wroc{\l}aw, Poland, e-mail: rolek@ift.uni.wroc.pl}\\
Physics Faculty and BiBoS, University of Bielefeld\\
D33615 Bielefeld, Germany}
\date{ }
\maketitle
\begin{abstract}
Departing from classical concepts of ergodic theory, formulated in terms of probability densities,
measures describing the mixing behavior and the loss of information in quantum open systems
are proposed. As application we discuss the chaotic outcomes of
continuous measurement processes in the EEQT framework. Simultaneous measurement of four
noncommuting spin components is shown to lead to a chaotic jumps on the quantum spin sphere
and to generate specific fractal images of a nonlinear iterated function system.
\end{abstract}

\vspace{1cm}
\newpage
\noindent
{\bf 1. Introduction}\\[4mm]
In the past two decades the study of chaotic dynamical systems has attracted attention of many physicists.
Roughly speaking, a system is said to be chaotic if orbits of the motion are in some sense irregularly
distributed. One of the most useful measures of this irregularity is the Kolmogorov-Sinai entropy. Another
important quantity relevant to control the
chaotic behavior is the Lyapunov exponents, which measure the exponential
instability of almost all orbits with respect to the change of initial conditions. It turns this instability
leads to the loss of memory of initial conditions, decay of correlations and approach to statistical
equilibrium.

It is believed that also quantum systems have different qualitative properties depending on whether the
corresponding classical systems are integrable or chaotic [1]. For example, they reveal significant differences
in the character of their wave functions or the distribution of their energy levels [2]. Also the numerical
analysis of some finite quantum systems shows quantitative differences in regular and chaotic regime [3].
It is obvious that on the quantum level
the distinction between chaotic and quasiperiodic behavior must be blurred, nevertheless some relative
measure of the degree of chaos should exist. There have been many attempts to find fingerprints of chaotic
behavior in quantum dynamical systems. The most natural one is based on
the correspondence principle, which states
that a quantum system is chaotic if its classical limit is chaotic. However,
it can only indicate some features of chaotic
behavior but cannot serve as a precise definition. As was pointed out by van Kampen [4] such notions like
ergodicity or mixing need a limit $t\to\infty$ while the correspondence principle
refers to $\hbar\to 0$, and these two limits
do not commute. Moreover, an example of a chaotic quantum phenomenon, which has no counterpart in the
classical limit was given [5]. One of the solutions of this obstacle is based on
the idea of taking the number of
degrees of freedom to infinity [6]. For example, in [7] it was shown that a class of quantum
dynamics of harmonic crystals becomes ergodic and mixing in the thermodynamic limit. Moreover, by taking
$\hbar\to 0$, classical properties of ergodicity and mixing are recovered. Similar results, but for the ideal
gas quantized according to the Maxwell-Boltzmann statistics, are presented in [8].

Another way to recognize chaotic behavior in quantum systems is to investigate the concept of entropy.
Kosloff and Rice [9] introduced a generalization of Kolmogorov-Sinai entropy for the quantum case, which
allows to compare the behavior of a given system when described alternatively by classical and quantum
mechanics. An even simpler idea was proposed by Thiele and Stone [10]. They suggested that the von Neumann
entropy of the time-averaged density matrix (minus the entropy connected with the preparation of the
initial state) can measure quantum chaos. The entropy point of view was taken also in a more recent paper by
S{\l}omczy\'{n}ski and \.{Z}yczkowski [11]. They considered a pair consisting of a quantum system and a
measuring apparatus and proposed a new definition of entropy to measure chaotic behavior of such a coupled
system. According to them ``the approach linking chaos with the unpredictability of the measurement outcomes
is the right one in the quantum case''.

On the other hand papers concerning the possible generalization of the notion of Lyapunov exponents have
appeared. Because there is no quantum analog of the classical trajectory, the starting point has to be
different. Perron-Frobenius operators acting on the space of densities provide a natural frame for the
construction of the quantum counterpart of classical characteristic exponents. Such a construction has been
carried out for example in [12,13].

In the present paper we will not discuss this interesting subject in general. Leaving aside the problem of
existence and definition of an intrinsic quantum chaos, we adopt the point of view of [11] and restrict
ourselves to the case of a quantum system interacting with a measuring device, or, more broadly, to quantum
open systems. Because, as was shown by Graham [14,15], in a class of such systems the sensitive dependence
of some expectation values on initial conditions remains, and the limits $t\to\infty$, $\hbar\to 0$ may
commute for a dissipative dynamics, so the notion of chaotic behavior seems to make sense there.
Moreover, in the presence of dissipation, coherence effects degrade and give way to an
incoherent dynamics closer to the classical behavior. One of the
attempts in this direction was the investigation of properties of the quantum survival probability function
in open systems [16]. For the following master equation
$$\dot{\rho}\;=\;-i[H,\,\rho ]\;+\;\gamma([H\rho,\,H]\:+\:[H,\,\rho H])$$
the different behavior in the regular and chaotic cases of the quantum survival probability function
averaged over initial conditions and Hamiltonian ensembles was demonstrated. For a general discussion
of quantum chaos with dissipation see [17]. A more general approach
investigating hypercyclicity and chaos in the context of strongly continuous semigroups of bounded linear
operators in Banach spaces was proposed in [18]. However, the definition of a chaotic semigroup given there
cannot be applied to contractive semigroups such as Perron-Frobenius semigroups acting on the space of
densities. In this case the idea of exactness of the system proved to be fruitful in the description of
chaotic Markov semigroups associated with some differential equations [19].

In this paper we generalize the notions of completely mixing and exact systems to the quantum level and
propose a quantity, the quantum characteristic exponent $\lambda_q$, which,
in the classical case, corresponds
to the highest order Lyapunov exponent measuring the speed of convergence of an exact classical system to
statistical equilibrium. The property $\lambda_q>0$ selects a subclass of completely mixing
systems which we call exponentially mixing. However, contrary to the
classical case, exponentially mixing quantum open systems may not imply chaotic behavior.
The relation of $\lambda_q$ to that one proposed by Majewski and Kuna in [13] is also
discussed. Finally, these concepts are illustrated by the examples of quantum measurements based on event
enhanced quantum theory (EEQT).\\[4mm]
{\bf 2. Classical systems}\\[4mm]
Classical mechanics deals with trajectories of dynamical systems whereas the time evolution in
quantum mechanics is formulated (in Schr\"odinger picture) in terms of density matrices. They correspond to
integrable positive functions in the commutative case. Therefore, to be closer to the framework of quantum
mechanics, we formulate some classical concepts of ergodic theory in terms of densities.

We start with recalling an intimate connection between the behavior of trajectories of a dynamical
system and the evolution of its densities. For a general discussion of this point see [19]. We
consider mainly the discrete time case. The reformulation of appropriate formulas to the case of
continuous time systems is straightforward. Suppose $\mbo$ is a normalized measure space ($\mu(X)\,=\,1$)
and $S:\,X\to X$ a measure preserving transformation of $X$, that is:\\
- $\forall A\in{\cal A}\quad S^{-1}(A)\in{\cal A}$,\\
- $\forall A\in{\cal A}\quad \mu(S^{-1}(A))\:=\:\mu(A)$.\\
Having defined $S$ we choose an initial point $x_0\in X$ and observe its trajectory $(x_0,\,S(x_0),\,
S^2(x_0),...)$. The chaotic behavior of the system can be recognized by a high sensitivity of the
trajectory with respect to a slight change of the initial state. It means that if we start with a
set of initial conditions located in a small region then, after a large number of iterations of the
transformation $S$, the points fill completely the whole space $X$. More precisely, for $S^n(A)\in{\cal A}$
\be \gran\mu(S^n(A))\;=\;1\quad\forall A\in{\cal A},\;\mu(A)\:>\:0 \ee
A transformation satisfying (1) is called {\bf exact}. Notice, that invertible transformations can not be exact.
A typical example of such a behavior is the r-adic transformation $S:\,[0,\,1)\to [0,\,1)$ given by
$S(x)\,=\,rx$(mod 1), $r\,=\,2,\,3,...$.

The behavior of trajectories can be described in terms of densities i.e. nonnegative, integrable
functions on $X$ with integral being equal to one. The space of densities will be denoted by $D$:
$$D\;=\;\{f\in\eli :\;f\geq 0\;a.e.,\;\|f\|_1\,=\,1\}$$
It is well known that densities and trajectories are related in the following sense.\\
{\bf Theorem 2.1} [19]. $S$ {\it is exact if and only if for all} $f\in D$
$$\gran\|P^nf\:-\:{\bf 1}\|_1\:=\;0,$$
{\it where $P$ is the Perron-Frobenius operator corresponding to the transformation $S$, $P:\,\eli\to\eli$,
given by}
$$ \int\limits_A(Pf)(x)d\mu(x)\;=\;\int\limits_{S^{-1}(A)}f(x)d\mu(x),\quad\forall A\in{\cal A} $$
For example, in the case of r-adic transformation and for the density $f(x)\,=\,2x$, we have
$$P^nf(x)\;=\;\frac{1}{r^n}\sumr 2(\frac{x\,+\,i}{r^n})\;=\;\frac{2x}{r^n}\;+\;\frac{r^n\,-\,1}{r^n}$$
and so $\|P^nf\:-\:{\bf 1}\|_1\,=\,\frac{1}{2r^n}$.\\
The concept of entropy gives a useful condition for a system to be exact. Let
\be H(f)\;=\;\calx\eta(f(x))d\mu(x), \ee
where $\eta(x)\,=\,-x\log x$ for $x>0$ and $\eta(0)\,=\,0$. Then
$\lim_{n\to\infty}H(P^nf)\,=\,0$ for all bounded $f\in D$ implies that system
$S$ is exact.
Observe that 0 is the maximum value of the entropy $H$. If the measure $\mu$ is finite but not
normalized, then the above theorem should be modified as follows. If
\be \gran H(P^nf)\;=\;H_{max}\;=\;H(\con) \ee
then $S$ is exact. If $\mu(X)\,=\,\infty$, then there is no constant density and we replace the
condition (3) by the following one
\be \gran H(P^nf|P^ng)\;=\;0\quad\mbox{for}\quad f,\,g\in D \ee
Here $H(f|g)$ is the relative entropy defined by
$$H(f|g)\;=\;\calx f(x)\log\frac{f(x)}{g(x)}d\mu(x)\quad\mbox{if supp}f\subset\mbox{supp}g$$
It has the following properties. $H(f|g)\geq 0$ and $H(Pf|Pg)\leq H(f|g)$.
Observe that, if there is a constant invariant density, then (4) implies (3). In fact,
putting $g\,=\,\con$ we get
$$H(P^nf|P^ng)\;=\;[-\:H(P^nf)\;+\;H(\con)]\to 0$$
We take conditions (3) or (4) as qualitative indicators of the chaotic behavior of dynamical
systems. In general, condition (4) gives less then (3) (see [20]):
$$\gran H(P^nf|P^ng)\;=\;0\;\Rightarrow\;\gran\|P^nf\:-\:P^ng\|_1\:=\;0$$
Such a system, in which $\|P^nf\:-\:P^ng\|_1\to 0$, $n\to\infty$, for any two densities, is called
{\bf completely mixing}. It is clear that every exact system is completely mixing. The converse is true
if there is an invariant density for the Perron-Frobenius operator $P$. However, the heat
equation generates a semigroup which is completely mixing but has no invariant density what shows
that the converse does not hold in general.

To describe quantitatively the chaotic behavior one can use the Lyapunov characteristic exponents.
The Lyapunov exponents of a given trajectory characterize the mean exponential rate of divergence
of nearby trajectories. In m-dimensional phase space there are m (possibly nondistinct) characteristic
exponents $$\sigma_1(x_o)\geq\sigma_2(x_0)\geq...\geq\sigma_m(x_o)$$
The largest one is defined by
$$\sigma_1(x_0)\;=\;\gran\limsup\limits_{d(x_1,x_0)\to 0}\frac{1}{n}\log\frac{d(S^nx_1,S^nx_0)}{d(x_1,x_0)},$$
where $d(x,\,y)$ is a metric on the phase space $X$. For example, in the r-adic case $\sigma(x_0)\,=\,\log r$
and does not depend on the initial point $x_0$.

Higher order exponents can be also defined [21]. The mean exponential growth of the phase space
volume given by
\be \sigma^{(m)}(x_0)\;=\;\lim\limits_{n\to\infty}\frac{1}{n}\log\frac{|S^nV(x_0)|}{|V(x_0)|}\;=\;
\sum\limits_{i=1}^m\sigma_i(x_0), \ee where $V(x_0)$ is a m-dimensional parallelepiped with one vertex placed in
$x_0$, is of particular interest. In view of theorem 2.1 the growth of the phase space volume corresponds to
decrease of any density to the constant density.
Therefore, for completely mixing systems we define another quantity
\be \lambda(f_0)\;=\;\liminf\limits_{\|f\,-\,f_0\|_1\to 0}
\gran[-\:\frac{1}{n}\log\frac{\|P^nf\:-\:P^nf_0\|_1}{\|f\:-\:f_0\|_1}] \ee
for $f_0,\,f\in D_0$, $D_0$ being some appropriately chosen dense subspace in $D$. Observe that for measure
preserving flows both $\sigma^m(x_0)$ and $\lambda(f_0)$ are equal to zero. We now discuss the elements of
the above formula more precisely. The minus sign reflects the property that for
fixed $f$ and $f_0$ $\|P^nf\,-\,P^nf_0\|_1\to 0$ when $n\to\infty$.
We have put the minus sign because our objective here is to replace
characteristic exponents, which are defined in terms of trajectories, by
a quantity expressed in terms of densities in such a way that they would
measure the same property of the given dynamics, and hence coincide for
simple one dimensional systems. As will be shown in proposition 2.2, in
the r-adic case the Lyapunov characteristic exponent $\sigma(x_0)\,=\,
\log r$, for any $x_0\in (0,\,1)$, indeed equals to our quantity $\lambda
({\bf 1})$. Therefore, although this minus sign may seem to be artificial,
it is necessary in order to describe the same feature of the dynamics.
Moreover, we use the liminf expression because for some densities, even
very regular ones, the action of the Perron-Frobenius operator may produce a constant density in a finite
number of iterations, what would result in an infinite limit. The restriction of $D$ to some subspace is a
crucial point. It is known from the analysis of mixing systems that the decay of correlations depends on
the functions chosen. As was shown in [22] in the case of a chaotic area preserving map on the
torus there are examples of functions for which the correlations decay faster than exponentially,
exponentially and only algebraically. Hence the rate of decay is sensitive to the choice of functions, and so
the restriction to a subset of ``nice'' functions is unavoidable. Finally, observe that formula (6) is
equivalent to $$ \lambda(f_0)\;=\;\inf\limits_{f\in D_0,f\neq f_0}\gran
[-\:\frac{1}{n}\log\frac{\|P^nf\:-\:P^nf_0\|_1}{\|f\:-\:f_0\|_1}] $$
In fact, if $f_k$ is an arbitrary sequence from $D_0$, then the sequence $g_k\,=\,f_0\,+\,\frac{f_k\,-\,f_0}{k}$
satisfies $\|g_k\,-\,f_0\|_1\to 0$ and
$$\frac{\|P^ng_k\:-\:P^nf_0\|_1}{\|g_k\:-\:f_0\|_1}\;=\;\frac{\|P^nf_k\:-\:P^nf_0\|_1}{\|f_k\:-\:f_0\|_1}$$
Moreover, because we calculate $n\to\infty$ limit first, the normalization can be dropped, and so
\be \lambda(f_0)\;=\;\inf\limits_{f\in D_0,f\neq f_0}\gran
[-\:\frac{1}{n}\log\|P^nf\:-\:P^nf_0\|_1] \ee
To show that formula (6) is indeed related with the Lyapunov exponents
we calculate $\lambda({\bf 1})$ in the r-adic case. At first we rescaled the interval
$[0,\,1)$ into $[0,\,2\pi)$, replace $dx$ by $dx/2\pi$ and define $S(x)\,=\,rx(\mbox{mod}2\pi)$. Hence $X$ can
be seen as the circle $S^1$.\\
{\bf Proposition 2.2}. {\it Let $D_0\,=\,D\cap C^1(S^1)$. Then} $\lambda({\bf 1})\,=\,\log r$.\\
{\bf Proof}: Let $\hat{f}$ denote the Fourier transform of a function $f\in D_0$, $f\neq f_0$, i.e.
$$\hat{f}\;=\;\frac{1}{2\pi}\int\limits_0^{2\pi}e^{-ikx}f(x)dx,\quad k\in{\bf Z}$$
Let $P$ be the Perron-Frobenius operator corresponding to $S$. Because
$$Pf(x)\;=\;\frac{1}{r}\sumi f(\frac{x}{r}\:+\:\frac{2\pi j}{r})$$ so
$$(Pf)^{\wedge}(k)\;=\;\frac{1}{2\pi r}\sumi\int\limits_0^{2\pi}e^{-ikx}f(\frac{x}{r}\:+\:\frac{2\pi j}{r})dx$$
$$=\;\frac{1}{2\pi}\sumi\int\limits_{\frac{2\pi j}{r}}^{\frac{2\pi(j+1)}{r}}e^{-ik(ry\,-\,2\pi j)}f(y)dy\;=\;
\hat{f}(kr)$$ and $(P^nf)^{\wedge}(k)\,=\,\hat{f}(kr^n)$. Hence
\be \|P^nf\:-\:{\bf 1}\|_{L^1}\:\leq\: \|P^nf\:-\:{\bf 1}\|_{L^2}\:=\;\|(P^nf)^{\wedge}\:-\:\hat{\bf 1}\|_{l^2}
\ee $$=\;[2\suki |\hat{f}(kr^n)|^2]^{1/2}$$
On the other hand \be \|f\:-\:{\bf 1}\|_{L^1}\:\geq\;\|\hat{f}\:-\:\hat{\bf 1}\|_{l^{\infty}}\:=\;\sup\limits_{
k\geq 1}|\hat{f}(k)| \ee
Because $\lim_{k\to\infty}|\hat{f}(k)|\,=\,0$, so $\|\hat{f}\:-\:\hat{\bf 1}\|_{l^{\infty}}\,=\,|\hat{f}(m)|$
for some natural $m$. Since $f'$ is continuous, hence integrable, so $\lim_{k\to\infty}k|\hat{f}(k)|\,=\,0$.
It implies that there exists $k_0$ such that for any $k\geq k_0$ we have $k|\hat{f}(k)|\leq |\hat{f}(m)|$.
Let us choose $n$ such that $r^n\geq k_0$. Then
\be [2\suki |\hat{f}(kr^n)|^2]^{1/2}\:\leq\;\frac{A|\hat{f}(m)|}{r^n},\quad\mbox{where}\quad A\;=\;[2\suki
\frac{1}{k^2}]^{1/2} \ee
Combining (8), (9) and (10) we arrive at
$$\frac{\|P^nf\:-\:{\bf 1}\|_{L^1}}{\|f\:-\:{\bf 1}\|_{L^1}}\;\leq\;\frac{A}{r^n}$$
Hence $\lambda({\bf 1})\geq\log r$. To finish the proof one only has to find a sequence of densities $f_k$
such that $\lim_{k\to\infty}\|f_k\,-\,{\bf 1}\|_1\,=\,0$ and
$$\lim\limits_{k\to\infty}\gran [-\:\frac{1}{n}\log\frac{\|P^nf_k\:-\:{\bf 1}\|_1}{\|f_k\:-\:{\bf 1}\|_1}]\;=\;
\log r$$ It may be easily checked that the sequence $f_k(x)\,=\,1\,+\,\frac{1}{k\pi}(x\,-\,\pi)$, $k\in{\bf N}$,
fulfills the required conditions. $\Box$\\
Completely mixing systems in which $\lambda (f)>0$ for some $f\in D$
will be called {\bf exponentially mixing}. As the example of r-adic
transformation shows they are closely related to chaotic systems.
Having discussed the signatures of the chaotic behavior in terms of
probability densities for classical systems
we now consider the quantum case.\\[4mm]
{\bf 3. Quantum open systems}\\[4mm]
As we mentioned in the introduction we discuss only dissipative quantum systems.
Dissipation in quantum theory appears by the coupling of a system to a reservoir.
The tracing over classical variables leads to the reduced dynamics.
Using certain approximation technics [23]
it can be shown that the reduced dynamics is given by a
Markov semigroup acting on the set of reduced density matrices.
The Hamiltonian evolution equation is replaced by the master equation. Such an evolution does not conserve
energy and maps pure states into mixed states. For technical reason we confine the discussion to N-level quantum
systems. Hence a dynamical semigroup $T_t:\,M_{N\times N}({\bf C})\to M_{N\times N}({\bf C})$ maps positive
definite matrices into positive definite matrices and preserves the trace $tr$. As the composition of a
Hamiltonian evolution and a conditional expectation $T_t$ is also completely positive.

As the first indicator of the mixing property we consider
the relative von Neumann entropy $H(\rho |\sigma)$
defined by $$H(\rho |\sigma)\;=\;tr(\rho\log\rho\;-\;\rho\log\sigma)$$ for any density matrices such that
supp$\rho\subset\,\mbox{supp}\sigma$.
It is well known that $H$ is non increasing with respect to $T_t$ i.e.
$H(T_t\rho |T_t\sigma)\leq H(\rho |\sigma)$. Following equation (4) we define a system to be completely mixing if
\be \liti H(T_t\rho |T_t\sigma)\;=\;0 \ee holds. If a totally mixed
state $\frac{1}{N}{\bf 1}$ is $T_t$-invariant, then (11) implies that $H(T_t\rho)\to H_{max}$. Since in the
classical case it leads to the exactness of a system we take this property as the definition of exactness
in the quantum case as well.
Let us notice that not every dissipative system is completely mixing. For example, for the
following master equation
$$\dot{\rho}\;=\;\sigma_1\rho\sigma_1\;-\;\rho$$ where $\rho$ is a $2\times 2$ density
matrix and $\sigma_1$ denotes the first Pauli matrix, the $t\to\infty$ limit of the relative entropy of two
neighboring one-dimensional projectors
$$e_1\;=\;\left(\begin{array}{cc} 1&0\\0&0\end{array}\right)\quad e_2\;=\;
\left(\begin{array}{cc} \cos^2\phi&\sin\phi\cos\phi\\ \sin\phi\cos\phi&\sin^2\phi\end{array}\right)$$
$\phi\in (0,\,\pi/4)$, is given by $\lim_{t\to\infty}H(T_te_1|T_te_2)\,=\,-\log\cos 2\phi\,\neq\,0$.
Let us also point out that Hamiltonian dynamics cannot be completely mixing since in this case the
relative entropy is constant in time.

Because $H(\rho|\sigma)\geq\frac{1}{2}\|\rho\,-\,\sigma\|_1^2$, where $\|A\|_1=\,tr|A|$, so
\be \liti H(T_t\rho|T_t\sigma)\;=\;0\Rightarrow\liti\|T_t\rho\:-\:T_t\sigma\|_1\;=\;0\ee
Therefore, for a completely mixing system one can ask the
question how fast the limit $\|T_t\rho\,-\,T_t\sigma\|_1,\,t\to\infty$, tends to zero.
Guided by the classical experience (compare formula (7)) we propose the
following quantity to measure the exponential rate of convergence
\be \lambda_q(\rho)\;=\;\inf\limits_{\sigma\neq\rho}\liti [-\frac{1}{t}\log\|T_t\rho\:-\:T_t\sigma\|_1]\ee
and call it the quantum characteristic exponent.
It is similar to that one proposed by Majewski and Kuna [13] (see also [24]).
Indeed, $\lambda_q(\rho)\,=\,\inf_{\sigma\neq\rho}\lambda_q(\rho;\,\sigma)$ and $\lambda_q(\rho;\,\sigma)$
coincides (up to the minus sign) with their formula if we notice that $T_t$ is linear and calculate the
corresponding derivative in a tangent direction to the space of all density
matrices, i.e. for $\delta x$ such that $tr\delta x\,=\,0$. As in the classical case completely mixing
systems with $\lambda_q(\rho)>0$, for some $\rho$, will be called exponentially mixing.
Finally, let us point out that if a completely mixing system has a
stationary density matrix $\rho_0$, that is $T_t(\rho_0)\,=\,\rho_0$ for
all $t\geq 0$, then $\rho_0$ is unique and $\lambda_q$ does not depend on
the choice of an initial statistical state $\rho$. In other words
$\lambda_q(\rho)\,=\,\lambda_q(\rho_0)$, where
$$\lambda_q(\rho_0)\;=\;\inf\limits_{\sigma\neq\rho_0}\lim\limits_{t\to
\infty}[-\frac{1}{t}\log\|T_t\sigma\:-\:\rho_0\|_1]$$

It is also worth noting that the entropy production rate for an evolving reduced density matrix of an open
quantum system proved to be a fruitful indicator of the character of dynamical behavior.
As was shown in [25] the
classical unpredictability corresponds to the rapid entropy production on the Lyapunov time-scale in
quantum analogs of classical systems which exhibit chaotic behavior. By contrast, dynamics of analogs of
integrable systems leads to a much slower evolution towards a dynamical equilibrium of the system.
Therefore, the rate of increase of entropy or, more generally, of decrease of relative entropy, can
indeed distinguish between chaotic and regular quantum evolutions, in a similar way as in the classical
case.\\[4mm]
{\bf 4. Continuous quantum measurements}\\[4mm]
EEQT is a minimal extension of quantum theory that accounts
for events [26,27]. It postulates that a quantum
measurement process is a particular coupling between a quantum and classical system. The time evolution of
such a hybrid system is determined by a completely positive semigroup of linear operators on the space of
density matrices of the total system. Moreover it provides the interpretation of the continuous evolution
of ensembles in terms of a piecewise deterministic process with values in the pure state space of the total
system. The process, after averaging,
reproduces the dynamical equation for statistical states. Such a process
turned out to be unique [28], what allows for deducing the algorithm generating sample histories of an
individual quantum system [29]. The sensitive dependence of sample paths of the associated process on the
initial conditions together with fractal structure of the limit set of a sample path indicate chaotic
properties of the corresponding dynamical semigroup. In this way they constitute the same counterpart for
the evolution of density matrices as trajectories do for density functions
in the classical case of $L^1(X)$ spaces.

Let us now briefly describe the framework of EEQT. Suppose that
possible states of the measuring apparatus $C$ form a discrete set
labeled by $\alpha\,=\,1,\,2,...,m$. The algebra of observables of
$C$ is the algebra ${\cal A}_c$ of complex finite sequences
$f_\alpha, \alpha =1,\ldots,m$. For technical reason we use
Hilbert space language even for the description of the classical
system. Thus we introduce an $m$-dimensional Hilbert space ${\cal
H}_c$ with a fixed basis, and we realize ${\cal A}_c$ as the
algebra of diagonal matrices $F=\dia(f_1,\ldots,f_m)$. Statistical
states of $C$ are then diagonal density matrices
$\dia(p_1,\ldots,p_m)$, and pure states of $C$ are vectors of the
fixed basis of ${\cal H}_c$. Suppose further $Q$ is the quantum
system whose bounded observables form the algebra $\ca_q$ of
bounded operators on a Hilbert space $\calh_q$. We assume that
$\calh_q$ is finite dimensional. Pure states of $Q$ form a complex
projective space ${\bf C}P(\calh_q)$ over $\calh_q$. Statistical
states of $Q$ are given by non-negative density matrices
${\hat\rho}$,  with $tr({\hat\rho})=1$. The algebra $\ca_T$ of
observables of the total system $T=Q\times C$ is given by the
tensor product of algebras of observables of $Q$ and $C$: $\ca_T
=\ca_q\otimes\ca_c$. It acts on the tensor product
$\calh_q\otimes\calh_c=\oplus_{\alpha=1}^m\calh_\alpha$, where
$\calh_\alpha\approx\calh_q.$ Thus $\ca_T$ can be thought of as
algebra of diagonal $m\times m$ matrices $A=(a_{\alpha\beta})$,
whose entries are quantum operators: $a_{\alpha\alpha}\in \ca_q$,
$a_{\alpha\beta}=0$ for $\alpha\neq\beta$. Statistical states of
$Q\times C$ are given by $m\times m$ diagonal matrices
$\rho=\dia(\rho_1,\ldots,\rho_m)$ whose entries are positive
operators on $\calh_q$, with the normalization
$tr(\rho)=\sum_\alpha tr(\ra)=1$. Duality between observables and
states is provided by the expectation value $<A>_\rho=\sum_\alpha
tr(\aaa\ra)$. The coupling of $Q$ to $C$ is specified by a matrix
$V=(\gab)$, where $\gab$ are linear operators: $\gab :\calh_\beta
\longrightarrow \calh_\alpha$. We assume  $g_{\alpha\alpha}=0$.
This condition expresses the simple fact: there is no dissipation
without receiving information. The evolution equation for states
is given by the Lindblad form \be {\dot
\rho}_\alpha=-i[H,\,\ra]+\sum_\beta \gab \rho_\beta \gab^\star -
{1\over2}\{\la,\ra\} \ee where $\la=\sum_\beta \gba^\star \gba$
and $\{\cdot,\cdot\}$ stands for the anticommutator. We apply now
the above scheme to three concrete measurement processes.\\ {\bf
4.1} {\it Measurement of noncommuting observables}. In this
example we model a simultaneous measurement of several
noncommuting observables, like different spin projections [30]. In
such a case we calculate the quantum exponent $\lambda_q$ for the
reduced dynamics given by tracing out classical parameters, and
show that it leads to a chaotic and fractal structure on the space
of pure states of the quantum system. An interesting investigation
of quantum chaos in open dynamical systems in which absorption
leads to the appearance of a fractal set in the underlying
classical phase space was presented in [31]. In the quantum case
fractal structure appears in the Husimi functions of eigenstates
of a non-unitary evolution operator. However, there is no apparent
relation between the approach taken in Ref. [31] and our
discussion of quantum fractal sets (on the set of pure states)
arising from a continuous simultaneous observation of several
noncommuting observables.

The measuring apparatus consists of four yes-no polarizers corresponding to spin directions $\vni$,
$i=\,1,...,4$, arranged at the vortices of a regular tetrahedron
$$\vec{n}_1\;=\;(1,\,0,\,0),\;\;\vec{n}_2\;=\;(-\frac{1}{3},\,0,\,\frac{2\sqrt{2}}{3})$$
$$\vec{n}_3\;=\;(-\frac{1}{3},\,\sqrt{\frac{2}{3}},\,-\frac{\sqrt{2}}{3}),\;\;
\vec{n}_4\;=\;(-\frac{1}{3},\,-\sqrt{\frac{2}{3}},\,-\frac{\sqrt{2}}{3})$$
Because the quantum system is a two-state system, so the quantum algebra is given by $2\times 2$ complex
matrices. We assume it evolves according to Hamiltonian $H\,=\,\frac{\omega}{2}\sigma_3$, $\omega\geq 0$,
where $\sigma_1,\,\sigma_2,\,\sigma_3$ denote Pauli matrices
$$\sigma_1\:=\:\left(\begin{array}{cc} 0& 1\\1& 0 \end{array}\right)\quad
\sigma_2\:=\:\left(\begin{array}{cc} 0& i\\-i& 0 \end{array}\right)\quad
\sigma_3\:=\:\left(\begin{array}{cc} -1& 0\\0& 1 \end{array}\right)$$
The coupling is specified by choosing four operators $a_i$ which correspond to four vectors $\vni$
\be a_i\;=\;\frac{1}{2}(I\:+\:\alpha\vni\cdot\vec{\sigma})\ee
where $\alpha\in [0,\,1]$. Notice, that for $\alpha\,=\,1$, $a_i$ are projection operators. The evolution
equation is given by
\be \dot{\rho}_i\;=\;-i[H,\,\rho_i]\;+\;\kappa\sum\limits_j a_i\rho_ja_i\;-\;\kappa a^2\rho_i\ee
what implies the following master equation for the reduced density matrix $\hat{\rho}\,=\sum_i\rho_i$
\be \dot{\hat{\rho}}\;=\;-i[H,\,\hat{\rho}]\;+\;\kappa\sum\limits_ia_i\hat{\rho}a_i\;-\;\kappa
a^2\hat{\rho}\ee
where $\kappa\geq 0$ is the coupling constant and $a^2\,=\,\sum_ia_i^2$. Eq.(17) implies that a projection
valued measure corresponding to a sharp measurement has been
replaced by a positive operator valued measure. Clearly, the totally mixed state $I/2$ is stationary.
At first we show that the dynamics given by eq.(17) leads to an exponentially mixing system.

To solve (17) we assume the initial state is a pure one i.e. $\hat{\rho}(0)\,=\,\frac{1}{2}(I\,+\,\vec{m}_0
\cdot\vec{\sigma})$, $\vec{m}_0\,=\,(c_1,\,c_2,\,c_3)$ with $c_1^2\,+\,c_2^2\,+\,c_3^2\,=\,1$. Because
the dynamical semigroup preserves positivity and trace so a
general solution is of the form $\hat{\rho}(t)\,
=\,\frac{1}{2}(I\,+\,\vec{m}(t)\cdot\vec{\sigma})$ with $\|\vec{m}(t)\|\leq 1$. By direct calculations we
obtain the following system of differential equations
$$\dot{m}_1\;=\;-\omega m_2\:-\:\kala m_1$$
$$\dot{m}_2\;=\;\omega m_1\:-\:\kala m_2$$
$$\dot{m}_3\;=\;-\kala m_3$$ and so
$$m_1(t)\;=\;(-c_2\sin\omega t\:+\:c_1\cos\omega t)\ekl$$
$$m_2(t)\;=\;(c_1\sin\omega t\:+\:c_2\cos\omega t)\ekl$$
$$m_3(t)\;=\;c_3\ekl$$
is the solution with the initial condition $\vec{m}(0)\,=\,\vec{m}_0$. Because
$$\|\hat{\rho}(t)\:-\:\frac{1}{2}I\|_1\;=\;\frac{1}{2}\|\vec{m}(t)\cdot\vec{\sigma}\|_1\;=\;\|\vec{m}(t)\|
\;=\;\ekl$$
and the convergence above does not depend on the choice of $\vec{m}_0$, so
\be \lambda_q(\frac{1}{2}I)\;=\;\inf\limits_{\hat{\rho}(0)\neq\frac{1}{2}I}\liti [-\frac{1}{t}\log\|
\hat{\rho}(t)\:-\:\frac{1}{2}I\|_1]\;=\;\kala\ee
We describe now a sample path of the process associated with the dynamical semigroup determined by eq.(17).
Assume that at time $t\,=\,0$ the quantum system is in the state $\vec{r}(0)\in S^2$ ( we identify here the
space of pure states of the quantum system with a two-dimensional sphere $S^2$ with radius 1). Under the
time evolution it evolves to the state $\vec{r}(t)$ which is given by the rotation of $\vec{r}(0)$ with
respect to z-axis. Then, at time $t_1$ a jump occurs. The time rate of jumps is governed by a homogeneous
Poisson process with rate $\kappa$. When jumping $\vec{r}(t)$ moves to
$$\vec{r}_i\;=\;\frac{(1\:-\:\alpha^2)\vec{r}(t)\:+\:2\alpha(1\:+\:\alpha\vec{r}(t)\cdot\vni)\vni}{1\:+\:
\alpha^2\:+\:2\alpha\vec{r}(t)\cdot\vni}$$
with probability $$p_i(\vec{r}(t))\;=\;\frac{1\:+\:
\alpha^2\:+\:2\alpha\vec{r}(t)\cdot\vni}{4(1\:+\:\alpha^2)}$$
And the process starts again. The iterations lead to a self-similar structure with sensitive dependence
on the initial state. Fig. 1, 2 and 3 depict sample paths of the quantum particle for different values of
$\alpha$ in the case when $H\,=\,0$ and $\kappa\,=\,1$. Fig. 4 a-c present $\alpha\,=\,0.7$ case together
with its zooms $\times 2$ and $\times 4$.\\
Numerical simulations performed for $H\,=\,0$ and $\kappa\,=\,1$ [32] show that
when $\alpha$ increases from 0.75 to 0.95, then the Hausdorff dimension of the limit set decreases from
1.44 to 0.49. It is worth noting that at the same time the quantum characteristic exponent $\lambda_q$
increases as $\alpha^2$ (see formula (18)). This suggests an intimate relation between $\lambda_q$ and the
Hausdorff dimension of the limit set. In the classical case
such a relation was established in [33,34].\\
{\bf 4.2} {\it Quantum Zeno effect}.
A phenomenon of keeping a quantum state from evolving by performing a sequence of frequent measurements was
discussed by Misra and Sudarshan [35] and named quantum Zeno effect (paradox). Roughly speaking it means that
the constant observation freeze the quantum system in its initial state. Such a situation was described, for
example, in [36] for the case of the neutron placed in a static magnetic field and interacting with a device
which selects one component of its spin. By this example we demonstrate a non-trivial dependence of
the quantum characteristic exponent $\lambda_q$ on the value of the coupling constant. Because the
quantum system is a two-state system, so the quantum algebra is just the algebra of $2\times 2$ complex
matrices, while the classical system being a yes-no device consists of two distinct points. The evolution
equation for the reduced density matrix
$\hat{\rho}\in M_{2\times 2}$, $\hat{\rho}\geq 0$, $tr\hat{\rho}\,=\,1$, is given by
\be \dot{\hat{\rho}}\;=\;-i[H,\,\hat{\rho}]\;+\;\kappa(e\hat{\rho}e\;-\;\frac{1}{2}\{e,\,\hat{\rho}\}),\ee
where Hamiltonian $H\,=\,\frac{\omega}{2}\sigma_3$ and the matrix $e\,=\,\frac{1}{2}(I\,+\,\sigma_1)$ is a
projector onto the first eigenvector of the Pauli matrix $\sigma_1$.
The coupling constant $\kappa$ measures the frequency of ``checking'' whether the quantum system
is in the eigenstate of $\sigma_1$ or not. Again, as in the previous example, the totally mixed state $I/2$
is stationary. Now eq.(19) leads to
the following system of differential equations
$$\dot{x}_1\;=\;-\omega x_2$$
\be \dot{x}_2\;=\;\omega x_1\;-\;\frac{\kappa}{2}x_2 \ee
$$\dot{x}_3\;=\;-\frac{\kappa}{2}x_3$$
where $\hat{\rho}(t)\,=\,\frac{1}{2}(I\:+\:\vec{x}\cdot\vec{\sigma})$. For a given initial condition
$\hat{\rho}(0)$ it is a routine exercise to solve (20), and hence to obtain the following result.\\
{\bf Theorem 4.1}. {\it Let $\kappa\,=\,4\alpha\omega$, $\alpha\geq 0$. Then}
$$\lambda_q(\frac{1}{2}I)\;=\;\left\{\begin{array}{ll} \omega\cdot\alpha&
\quad\mbox{if}\quad\alpha\in [0,\,1]\\
\omega\cdot\frac{1}{\alpha\:+\:\sqrt{\alpha^2-1}}&\quad\mbox{if}\quad\alpha>1\end{array}\right.$$
It follows that the degree of mixing (measured in terms of $\lambda_q$) increases from zero to the
maximum value $\omega$ when the frequency of measurements increases to the critical value $4\omega$.
The further increase of number of observations
results in the decrease of $\lambda_q$ to 0 like $\frac{1}{\kappa}$.
Hence we can conclude that the interaction with the measuring
apparatus makes the evolution of the quantum system
exact (and so mixing), but constant observation suppresses this property. \\
{\bf 4.3} {\it Two-level atom driven by a laser}.
Let us consider the fluorescent photons emitted by a
single, two-level atom that is coherently driven by an external
electromagnetic field. It is known that the quantum system evolves from
the ground state in a dissipative way. When a photoelectric count is
recorded by a photoelectric detector (we assume the detector efficiency
to be equal to one), the atom returns to the ground state with the emission of one photon.
The master equation for such a system is given by the following formula
\be\dot{\rho}\;=\;-i[H,\,\rho ]\;+\;\gamma A\rho A^*\;-\;\frac{\gamma}{2}\{A^*A,\,\rho\}\ee
where $$H\;=\;-\frac{\Omega}{2}\sigma_1\quad\mbox{and}\quad
A\;=\;\left(\begin{array}{cc}0&0\\1&0\end{array}\right)$$
$\Omega$ is Rabi frequency and $\gamma>0$ is the relaxation rate.
Because the unique stationary state $\rho_0$ is given by $\rho_0\,=\,\frac{1}{2}(I\,+\,n_2\sigma_2\,+\,
n_3\sigma_3)$, where
$$n_2\;=\;\frac{2\Omega\gamma}{4\omega^2+\gamma^2}\qquad n_3\;=\;\frac{-\gamma^2}{4\Omega^2+\gamma^2}$$
so the system is not exact. However, it is exponentially mixing. Solving eq.(21) and using formula (13)
we obtain that $\lambda_q(\rho_0)\,=\,\gamma/2$. Hence, although $\lambda_q(\rho_0)>0$ it would be
difficult to consider this system chaotic. The property of being exponentially mixing is a new feature
in quantum open systems.\\[4mm]
{\bf Acknowledgment}\\
We would like to thank the referees for their remarks which helped us to improve the clarity of the paper.
One of the authors (R.O.) would like to thank A. von Humboldt Foundation for the financial support.\\[4mm]
{\bf References}\\[4mm]
$[1]$ Casati G., Quantum mechanics and chaos, in: Chaos and Quantum Physics, eds. M. J. Gianoni, A. Voros
and J. Zinn-Justin, North Holland, 1991\\
$[2]$ Gutzwiller M. C., Chaos in Classical and Quantum Mechanics, Springer, New York, 1990\\
$[3]$ Alicki R., Makowiec D., Miklaszewski W., Quantum chaos in terms of entropy for periodically kicked top,
Phys. Rev. Lett. {\bf 75} (1996) 838-841\\
$[4]$ van Kampen N. G., Quantum chaos as basis for statistical mechanics, in: Chaotic Behaviour in
Quantum Systems, ed. G. Casati, Plenum Press, New York, 1985\\
$[5]$ Jona-Lasinio G., Presilla C., Capasso F., Chaotic quantum phenomena without classical counterpart,
Phys. Rev. Lett. {\bf 68} (1992) 2269-2272\\
$[6]$ Jona-Lasinio G., Presilla C., Chaotic properties of quantum many-body systems in the thermodynamic
limit, Phys. Rev. Lett. {\bf 77} (1996) 4322-4325\\
$[7]$ Graffi S., and Martinez A.,
Ergodic properties of infinite quantum harmonic crystals: an analytic approach,
J. Math. Phys. {\bf 37} (1996) 5111-5135\\
$[8]$ Lenci M., Ergodic properties of the quantum ideal gas in the Maxwell-Boltzmann statistics,
J. Math. Phys. {\bf 37} (1996) 5136-5157\\
$[9]$ Kosloff R., Rice S. A., The influence of quantization on the onset of chaos in Hamiltonian systems:
the Kolmogorov entropy interpretation, J. Chem. Phys. {\bf 74} (1981) 1340-1349\\
$[10]$ Thiele E., Stone J., A measure of quantum chaos, J. Chem. Phys. {\bf 80} (1984) 5187-5193\\
$[11]$ S{\l}omczy\'{n}ski W., \.{Z}yczkowski K., Quantum chaos: an entropy approach, J. Math. Phys.
{\bf 35} (1994) 5674-5700\\
$[12]$ Vilela Mendez R., Entropy and quantum characteristic exponents. Steps towards a quantum Pesin theory,
in: Chaos - The Interplay Between Stochastic and Deterministic Behaviour, eds. P. Garbaczewski, M. Wolf
and A. Weron, Springer, Berlin, 1995\\
$[13]$ Majewski W. A., Kuna M., On quantum characteristic exponents, J. Math. Phys.
{\bf 34} (1993) 5007-5015\\
$[14]$ Graham R., Chaos in dissipative quantum systems, in: Chaotic Behaviour in Quantum Systems, ed.
G. Casati, Plenum Press, New York, 1985\\
$[15]$ Graham R., Chaos in dissipative quantum maps, in: Quantum Measurement and Chaos, eds. E. R. Pike
and S. Sarkar, Plenum Press, New York, 1987\\
$[16]$ Tameshtit A., Sipe J. E., Survival probability and chaos in an open system, Phys. Rev. A
{\bf 45} (1992) 8280-8283\\
$[17]$ Dittrich T., H\"anggi P., Ingold G-L., Kramer B., Sch\"on G., Zwerger W., Quantum Transport and
Dissipation, Wiley-VCH, Weinheim, 1998\\
$[18]$ Desch W., Schappacher W., Webb G. F., Hypercyclic and chaotic semigroups of linear operators,
Ergod. Th. \& Dynam. Systems {\bf 17} (1997) 793-819\\
$[19]$ Lasota A., Mackey M. C., Chaos, Fractals and Noise. Stochastic Aspects of Dynamics,  2nd ed
Springer, New York, 1994\\
$[20]$ {\L}oskot K., Rudnicki R., Relative entropy and stability of stochastic semigroups,
Ann. Pol. Math. {\bf 53} (1991) 139-145\\
$[21]$ Lichtenberg A. J., Lieberman M. A., Regular and Stochastic Motion, Springer, New York, 1983\\
$[22]$ Crawford J. D., Cary J. R., Decay of correlations in a chaotic measure-preserving transformations,
Physica D {\bf 6} (1983) 223-232\\
$[23]$ Alicki R., Lendi K., Quantum Dynamical Semigroups and Applications,
Lect. Notes Phys. {\bf 286}, 1987\\
$[24]$ Majewski W. A., Remarks on quantum characteristic exponents, Chaos, Solitons and Fractals {\bf 9}
(1998) 77-82\\
$[25]$ Zurek W. H., Paz J. P., Quantum chaos: a decoherent definition, Physica D {\bf 83} (1995) 300-308\\
$[26]$ Blanchard Ph., Jadczyk A., On the interaction between classical and quantum systems, Phys. Lett. A
{\bf 175} (1993) 157-164\\
$[27]$ Blanchard Ph., Jadczyk A., Strongly coupled quantum and classical systems and Zeno's effect,
Phys. Lett. A {\bf 183} (1993) 272-276\\
$[28]$ Jadczyk A., Kondrat G., Olkiewicz R., On
uniqueness of the jump process in event enhanced quantum theory, J. Phys. A {\bf 30} (1997) 1863-1880\\
$[29]$ Blanchard Ph., Jadczyk A., Event enhanced quantum
theory and piecewise deterministic dynamics, Ann. der Physik {\bf 4} (1995) 583-599\\
$[30]$ Jadczyk A., Topics in quantum dynamics, in: Infinite Dimensional
Geometry, Noncommutative Geometry, Operator Algebras and Fundamental Interactions,
Coquereaux R et al. (eds.), World Scientific, Singapore, 1995\\
$[31]$ Casati G., Maspero G., Shepelyansky D.L., Quantum fractal eigenstates, Physica D {\bf 131} (1999)
311-316\\
$[32]$ Jastrzebski G., Interacting classical and quantum systems. Chaos from
quantum measurements, Ph. D. thesis, University of Wroc{\l}aw (in Polish), 1996\\
$[33]$ Manning A., A relation between Lyapunov exponents, Hausdorff dimension and entropy,
Ergod. Th. \& Dynam. Systems {\bf 1} (1981) 451-459\\
$[34]$ Young L.-S., Dimension, entropy and Lyapunov exponents
Ergod. Th. \& Dynam. Systems {\bf 2} (1982) 109-124\\
$[35]$ Misra B., Sudarshan E. C. G., The Zeno's paradox in quantum theory, J. Math. Phys.
{\bf 18} (1977) 756-763\\
$[36]$ Pascazio S., Namiki M., Badurek G., Rauch H., Quantum Zeno effect with neutron spin, Phys. Lett. A
{\bf 179} (1993) 155-160\\
\end{document}